\documentclass[aps,prd,reprint,twocolumn,superscriptaddress,showpacs,notitlepage]{revtex4-1}

\usepackage{graphicx}
\usepackage{mathrsfs}
\usepackage{bm}
\usepackage{amsmath}
\usepackage{dcolumn}
\usepackage{epstopdf}
\usepackage{dsfont}
\usepackage{amssymb}
\usepackage{tabularx}
\usepackage{array}
\usepackage{float}
\usepackage{color}
\usepackage{epstopdf}
\usepackage{mathrsfs}
\usepackage[colorlinks, linkcolor=blue,anchorcolor=blue,citecolor=blue,urlcolor=blue]{hyperref}
\usepackage[T1]{fontenc}

\begin{document}
\title{Phononic real Chern insulator with protected corner modes in graphynes}

\author{Jiaojiao Zhu}
\address{Research Laboratory for Quantum Materials, Singapore University of Technology and Design, Singapore 487372, Singapore}

\author{Weikang Wu}\email{weikang.wu@ntu.edu.sg}
\address{Division of Physics and Applied Physics, School of Physical and Mathematical Sciences, Nanyang Technological University, Singapore, 637371, Singapore}
\address{Research Laboratory for Quantum Materials, Singapore University of Technology and Design, Singapore 487372, Singapore}

\author{Jianzhou Zhao}\email{jzzhao@swust.edu.cn}
\address{Co-Innovation Center for New Energetic Materials, Southwest University of Science and Technology, Mianyang 621010, China}
\address{Research Laboratory for Quantum Materials, Singapore University of Technology and Design, Singapore 487372, Singapore}

\author{Cong Chen}
\address{School of Physics, and Key Laboratory of Micro-nano Measurement-Manipulation and Physics, Beihang University, Beijing 100191, China}

\author{Qianqian Wang}
\address{Research Laboratory for Quantum Materials, Singapore University of Technology and Design, Singapore 487372, Singapore}

\author{Xian-Lei Sheng}
\address{School of Physics, and Key Laboratory of Micro-nano Measurement-Manipulation and Physics, Beihang University, Beijing 100191, China}

\author{Lifa Zhang}
\affiliation{NNU-SULI Thermal Energy Research Center (NSTER) \& Center for Quantum Transport and Thermal Energy Science (CQTES), School of Physics and Technology, Nanjing Normal University, Nanjing 210023, China}

\author{Y. X. Zhao}\email{zhaoyx@nju.edu.cn}
\affiliation{National Laboratory of Solid State Microstructures and Department of Physics, Nanjing University, Nanjing 210093, China}
\affiliation{Collaborative Innovation Center of Advanced Microstructures, Nanjing University, Nanjing 210093, China}

\author{Shengyuan A. Yang}
\address{Research Laboratory for Quantum Materials, Singapore University of Technology and Design, Singapore 487372, Singapore}


\begin{abstract}

Higher-order topological insulators have attracted great research interest recently. Different from conventional topological insulators, higher-order topological insulators do not necessarily require spin-orbit coupling, which makes it possible to realize them in spinless systems.
Here, we study phonons in 2D graphyne family materials. By using first-principle calculations and topology/symmetry analysis, we find that phonons in both graphdiyne and $\gamma$-graphyne exhibit a second-order topology, which belongs to the specific case known as real Chern insulator. We identify the nontrivial phononic band gaps, which are characterized by nontrivial real Chern numbers enabled by the spacetime inversion symmetry. The protected phonon corner modes are verified by the calculation on a finite-size nanodisk.
Our study extends the scope of higher-order topology to phonons in real materials. The spatially localized phonon modes could be useful for novel phononic applications.

\end{abstract}

\maketitle

\section{Introduction}
The study of topological insulators has developed into a vast research field~\cite{insulator-1,insulator-2,insulator-3,insulator-4}. A conventional $d$-dimensional topological insulator features a nontrivial bulk topological invariant defined for the valence states below the bulk band gap, and it has protected gapless states at its $(d-1)$-dimensional boundaries. Later on, it was realized that with certain spatial symmetries, there could exist a class of higher-order topological insulators (HOTIs), in which the gapless states appear not at $(d-1)$- but at $(d-n)$-dimensional boundaries with $n>1$~\cite{hoti-1,hoti-2,hoti-3,hoti-4, hoti-5, hoti-6}. For example, a two-dimensional (2D) second-order topological insulator would have gapped bulk and edge spectra, meanwhile, gapless excitations occur at its 0D corners.
The concept of HOTI has attracted great interest~\cite{hoti-interest-1,hoti-interest-2,hoti-interest-3,hoti-interest-4, hoti-interest-6, hoti-interest-5}. Unlike the conventional topological insulator, a HOTI does not necessarily require the spin-orbit coupling. In other words, HOTI can be realized in spinless systems. Thus, its impact is not limited to electronic systems, but also spreads into bosonic and even classical systems, such as photonic/acoustic metamaterials~\cite{hoti-photonic-1,hoti-photonic-2,hoti-acoustic-1,hoti-acoustic-2,hoti-meta}, electric circuit arrays~\cite{hoti-electric-circuit-1, hoti-electric-circuit-2}, and mechanical networks~\cite{hoti-mechanical-1, hoti-mechanical-2}.

Phonons are the fundamental bosonic excitations in a crystal material, describing the collective vibrations of atomic lattice. Phonons play a crucial role in the material's thermal properties, interact closely with other quasiparticles or collective excitations, such as electrons, photons, magnons, etc., and underly remarkable effects such as superconductivity. Recently, there is a surge of interest in transferring the topological physics to phonons~\cite{phonon-7, phonon-8, phonon-1, phonon-3, phonon-2, phonon-4, phonon-5, phonon-9, phonon-11, phonon-12, phonon-6, zhu2021symmetry}. However, so far, most studies are focusing on the gapless phonon states, such as nodal points and nodal lines in the phonon spectrum, which are analogues of topological semimetals in the electronic context. Meanwhile, gapped topological phonon states receive less attention. Particularly, to our knowledge, there is no proposal of a realistic material which exhibits a HOTI state in phonons.

In this work, we fill this gap by showing that phonons in the 2D graphyne family materials exhibit the second-order topology. This study is motivated by our recent proposal of these materials as the first example of 2D electronic HOTIs~\cite{hoti-interest-6, 2d-ehoti-2}. We have shown that their HOTI state belongs to the specific case known as real Chern insulator~\cite{real-chern-insulator}, which is characterized by a nontrivial real Chern number $\nu_R$ and enabled by the spacetime inversion symmetry $\mathcal{PT}$. Here, we show that the discussion can be naturally extended to phonons. Phonons are spinless, obey bosonic statistics, and respect the same crystal symmetries including $\mathcal{PT}$. Hence, each phononic band gap in the spectrum can be characterized by a real Chern number $\nu_R\in\mathbb{Z}_2$. When $\nu_R=1$, it indicates that the corresponding phononic gap is topologically nontrivial, and there will be protected phonon modes localized at corners for a finite sample that preserves $\mathcal{PT}$. Taking graphdiyne and graphyne from the family as examples and using first-principles calculations, we compute the real Chern numbers for their phononic band gaps via the parity eigenvalue approach.
The phonon corner modes are explicitly verified by the calculation on a disk geometry. Our approach can be readily applied to other materials to reveal possible higher-order topology in phonons. In addition to the fundamental interest, the results here also open a new route to realize localized phonons, which might be useful for novel device applications.

\section{Calculation Method}
Density functional theory (DFT) calculations were conducted by using the Vienna ab initio simulation package (VASP)~\cite{vasp-1, vasp-2}. The projector augmented wave (PAW) pseudopotentials were adopted in the calculation~\cite{paw-1, paw-2}. Generalized gradient approximation (GGA) in the form of Perdew-Burke-Ernzerhof (PBE) realization was adopted for the exchange-correlation potential~\cite{pbe}. The valence electrons treated in the calculations include C (2s$^2$2p$^2$). The kinetic energy cutoff was fixed to 520 eV. $\Gamma$-centered $10 \times 10 \times 1$ $k$ point mesh was adopted for the self-consistent calculations. The energy and force convergence criteria were set to be 10$^{-7}$ eV and 0.001 $\text{eV/\AA}$, respectively. We used density functional perturbation theory (DFPT)~\cite{dfpt} in combination with the Phonopy package~\cite{phonopy} to obtain the force constants and phonon spectra.A supercell of $3 \times 3 \times 1$ is adopted for the calculation of force constants. For computing the phonons for the nanodisk geometry, we first calculated the second rank tensor of force constants in Cartesian coordinates from DFPT, from which we extracted the parameters for constructing the phononic tight-binding model. The spatial distribution of phonon modes was calculated by using Pybinding package~\cite{Moldovan2020pybinding}.

\section{Structure and phonon spectrum}

Graphynes represent a family of 2D carbon allotropes composed of $sp$ and $sp^2$ hybridized carbon atoms~\cite{1987, carbon-1, carbon-3, carbon-2}. They can be derived from the graphene structure by inserting acetylenic linkages in different manners. Their structural models were first proposed by Baughman \emph{et al.} in 1987~\cite{1987}. Some members of this family have been successfully realized in experiment. The most prominent example is graphdiyne, which was first synthesized in 2010 by Li \emph{et al.} via a cross-coupling reaction method~\cite{carbon-3}. Meanwhile, as the representative of the family, $\gamma$-graphyne (often simply referred to as graphyne) has been realized in experiment in the form of small fragments~\cite{carbon-1, fragments}. Hence, in this work, we will focus on these two examples. We will first discuss the results of graphdiyne in the following sections. The results for $\gamma$-graphyne are similar and will be briefly summarized in Sec.~\ref{gyn}.

\begin{figure}[tb!]
    \centering
    \includegraphics[width=0.9\linewidth]{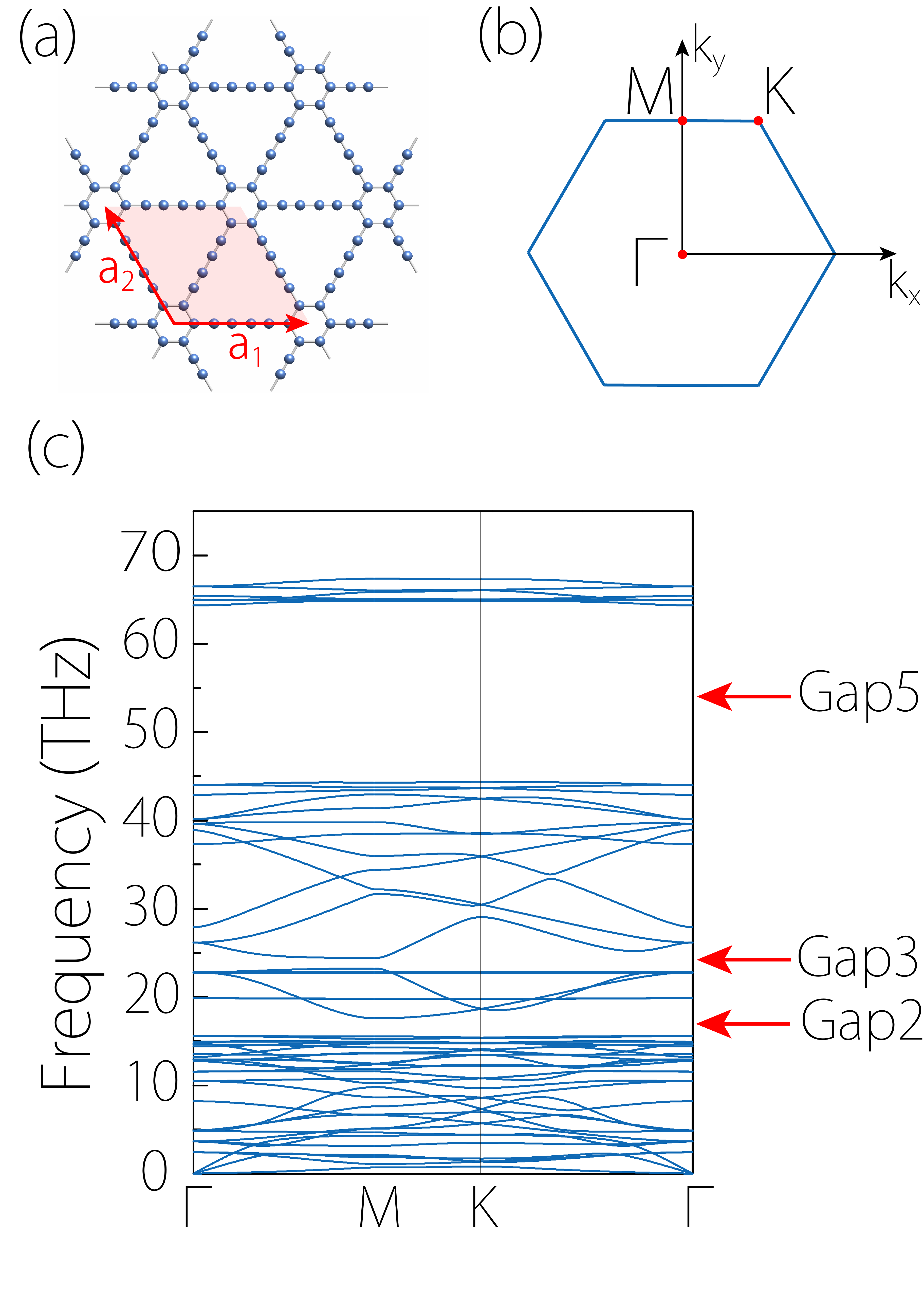}
    \caption{(a) Crystal structure of graphdiyne. The shaded region indicates the unit cell. (b) Brillouin zone with high symmetry points labeled. (c) Calculated phonon spectrum of graphdiyne. The arrows indicate three of the five global phononic band gaps. Other two gaps are small hence are not indicated here. }
    \label{fig:1}
\end{figure}

The structure of graphdiyne is illustrated in Fig.~\ref{fig:1}(a). The lattice is completely flat with single-atom thickness. It can be viewed as formed by inserting one diacetylenic linkage between two neighboring benzene rings in the graphene structure while maintaining its $p6m$ symmetry. Graphdiyne exhibits a high $\pi$ conjugation, which helps to stabilize the 2D planar structure and lower the energy. In fact, it was shown that graphdiyne is the most stable among 2D carbon allotropes that contain diacetylenic linkages~\cite{stable}. Each unit cell of the structure contains 18 carbon atoms. The optimized lattice
constant from our first-principles calculations is 9.460 $\text{\AA}$, which is consistent with previous results.

The calculated phonon spectrum of graphdiyne is plotted in Fig.~\ref{fig:1}(c). One observes that the spectrum spans a range up to about 70 THz, which is comparable to that of graphene ($\sim 50$ THz), implying its relatively strong bonding.
There are totally 54 phonon branches (3 acoustic branches plus 51 optical branches), corresponding to the 18 atoms in a unit cell. Among the three acoustic branches, two have a linear dispersion approaching the $\Gamma$ point, whereas the remaining out-of-plane (ZA) acoustic branch has a quadratic dispersion, which is a characteristic feature of 2D materials~\cite{feature-1, feature-2}. The sound speed for the longitudinal acoustic phonons is about 18.2 km/s, which is much larger than that of the MoS$_2$ ($\sim$ 6.5 km/s)~\cite{mos2}, and close to graphene ($\sim$ 21.2 km/s)~\cite{graphene}.

Notably, we can observe several phononic band gaps in the spectrum, which are labeled from Gap 1 to Gap 5 in Fig.~\ref{fig:1}(c) with increasing energy. Each gap separates the whole phonon spectrum into two groups: the phonon bands below the gap and the bands above, analogous to the valence and conduction bands in the electronic context. Thus, for each gap, we can define a real Chern number based on the bands below the gap (i.e., the ``valence'' bands) to characterize the topology of the gap.

Before proceeding, we should comment on two important differences between phonons and electrons. First, electrons are fermions, whereas phonons are bosons. They follow distinct statistics. It follows that for electronic systems, the only meaningful band gap to analyze is the one that near the Fermi energy; but for phonons, all the band gaps are meaningful to discuss. Second, the phonon energy scale (few tens of THz) is much smaller than the electron's. This makes the whole phonon spectrum detectable in experiment, e.g., by optical or neutron scattering. For example, it is demonstrated in recent experiment with inelastic x-ray scattering technique, which can achieve the meV-resolution~\cite{mev-1, mev-2, mev-3}. This means that all the phononic band gaps are also experimentally accessible.

The graphdiyne system preserves the $\mathcal{PT}$ symmetry. $\mathcal{PT}$ operates locally at every $k$ point of the Brillouin zone (BZ). For a spinless system like phonons, we have $(\mathcal{PT})^2=1$. Up to a possible unitary transformation, it can always be represented as $\mathcal{PT}=\mathcal{K}$, where $\mathcal{K}$ is the complex conjugation. This means that with $\mathcal{PT}$ symmetry, all the phonon eigenmodes $u_n(\bm k)$ can be made real, i.e., $u_n(\bm k)=u^*_n(\bm k)$. In Ref.~\cite{real-chern-insulator}, it has been shown that the real bands in 2D are characterized by a $\mathbb{Z}_2$ invariant, the real Chern number $\nu_R$, which is analogous to the conventional Chern number for complex bands.
When there are only two bands below the gap, $\nu_R$ has a integral expression in terms of the real Berry curvature of these bands, as shown in Ref.~\cite{real-chern-insulator}.  However, when there are more than two bands, an integral expression for $\nu_R$ is not available.
Nevertheless, for systems that also preserve the inversion symmetry $\mathcal{P}$ (which is the case here), $\nu_R$ can be evaluated by using the parity analysis approach~\cite{Ahn2018}.

In this approach, one needs to analyze the parity eigenvalues of phonon modes at the four inversion-invariant momentum points $\Gamma_i$ $(i=1,\cdots, 4)$ of the BZ, which include $\Gamma$ and the three $M$ points [see Fig.~\ref{fig:1}(b)]. For each phononic band gap, at point $\Gamma_i$, one counts the number of phonon bands below the gap that have the negative parity eigenvalue. Denote this number by $n^{\Gamma_i}_-$. Then, $\nu_R$ can be obtained as~\cite{Ahn2018}
\begin{equation}\label{real}
   (-1)^{\nu_R} = \prod_{i=1}^{4} (-1)^{\left\lfloor (n_{-}^{\Gamma_{i}}/2) \right\rfloor}.
\end{equation}
where $\left\lfloor \cdots \right\rfloor$ is the floor function.

We apply this approach to analyze the 5 phononic band gaps in graphdiyne. The obtained results are listed in Table ~\ref{123}. One can see that Gap 1 and Gap 2 have a nontrivial real Chern number $\nu_R=1$, so these gaps correspond to a phononic real Chern insulator state.

\begin{table*}
\renewcommand\arraystretch{1.5}
\caption{\label{123} Topological characterization of the phononic band gaps in graphdiyne. Here, $N$ is the number of bands below the gap. $n_-^\Gamma$ and $n_-^M$ are the number of bands (out of the $N$ bands) with negative parity eigenvalues at $\Gamma$ and $M$, respectively. $\nu_R$ gives the corresponding real Chern number.}
\begin{tabular}{cccccc}
\hline\hline
Gap No.  &\qquad Frequency range (THz)& \qquad Bands below the gap ($N$) & \qquad $n_{-}^{\Gamma}$  & \qquad $n_{-}^{M}$ & \qquad$\nu_R$ \\
\hline
1 & \qquad $15.0 - 15.2$ & \qquad 28 & \qquad  16 & \qquad 14 & \qquad 1\\
2 & \qquad $15.6 - 17.6$ & \qquad 30 & \qquad  16 & \qquad 14 & \qquad 1\\
3 & \qquad $23.2 - 24.4$ & \qquad 36 & \qquad  18 & \qquad 18 & \qquad 0\\
4 & \qquad $39.8 - 40.1$ & \qquad 43 & \qquad  22 & \qquad 22 & \qquad 0\\
5 & \qquad $44.2 - 64.4$ & \qquad 48 & \qquad  24 & \qquad 24 & \qquad 0\\
\hline\hline
\end{tabular}
\end{table*}

Let's consider Gap 2, which is the gap between 15.6 THz and 17.6 THz. Below this gap, there are 30 phonon branches. At $\Gamma$, there are 14 branches with positive parity eigenvalues and 16 branches with negative parity eigenvalues, i.e., $n^\Gamma_+=14$ and $n^\Gamma_-=16$. In comparison, at the three $M$ points, we find $n^M_+=16$ and $n^M_-=14$. This mismatch signals a double band inversion between $\Gamma$ and $M$, which is captured by the nontrivial real Chern number. Meanwhile, Gap 1 has a smaller size from 15.00 THz to 15.25 THz. The analysis is similar, so we will not elaborate here.

\section{Corner phonon mode}

It has been shown that a real Chern insulator has a second-order topology~\cite{2d-ehoti-2}, i.e., there must exist 0D corner modes for any sample that preserves the $\mathcal{PT}$ symmetry. Since we have identified phononic band gaps in graphdiyne that feature nontrivial real Chern numbers, it means that in a graphdiyne nanodisk, we should have localized corner phonon modes in these gaps.

\begin{figure}[ht!]
    \centering
    \includegraphics[width=1.0\linewidth]{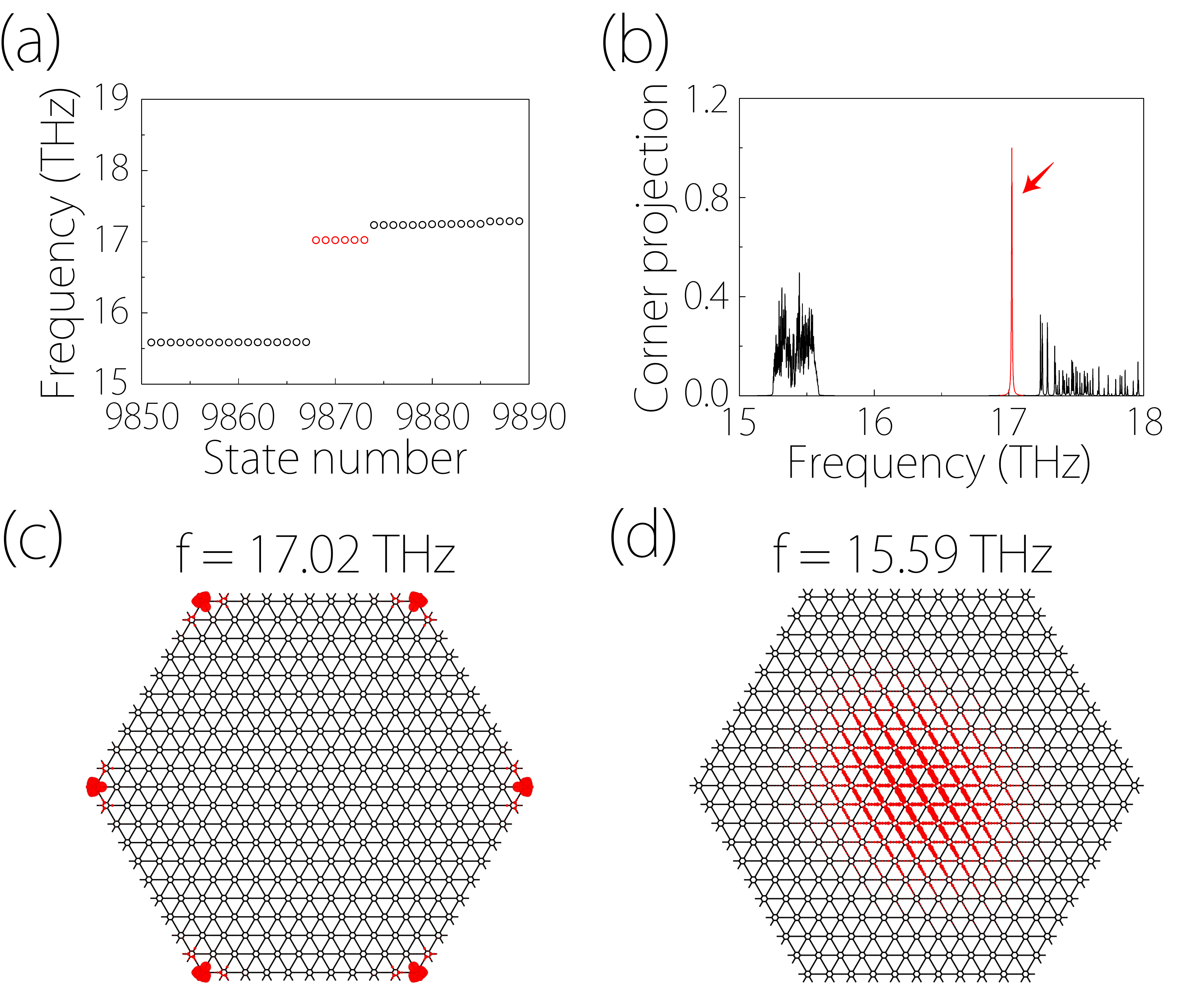}
    \caption{(a) Phonon spectrum for the graphdiyne nanodisk in (c). Here, we focus on the part of frequency range close to the bulk Gap 2. (b) Corner projection spectrum, which is the phonon local density of states at one corner of the disk. The arrow indicates the peak due to the corner modes. (c) Spatial distribution of the phonon modes at 17.02 THz, i.e. the red colored modes inside Gap 2 in (a), which are localized at the corners of the disk. For comparison, (d) shows the distribution of phonon modes at a nearby frequencies (15.59 THz), which is extended in the bulk of the disk.}
    \label{fig:2}
\end{figure}

To confirm this prediction, we calculate the phonon spectrum for the graphdiyne nanodisk geometry. The calculation is based on an \emph{ab initio} tight binding model extracted from the first-principles calculations. In Fig.~\ref{fig:2}(a), we plot the phonon spectrum close to Gap 2. One observes that at 17.02 THz inside Gap 2, there are 6 degenerate phonon modes. We have checked the spatial distribution of these modes and found that they are localized at the six corners of the disk, as illustrated in Fig.~\ref{fig:2}(c). As a comparison, in Fig.~\ref{fig:2}(d), we also plot the spatial distribution of the nearby modes at 15.59 THz, and one can easily see that the conventional modes is not localized but spread throughout the sample. In Fig.~\ref{fig:2}(b), we plot the local density of states at one corner (taking the mode weight within one unit cell of the corner). One can clearly observe the sharp peak due to the localized corner modes. 

\begin{figure}[ht!]
    \centering
    \includegraphics[width=1.0\linewidth]{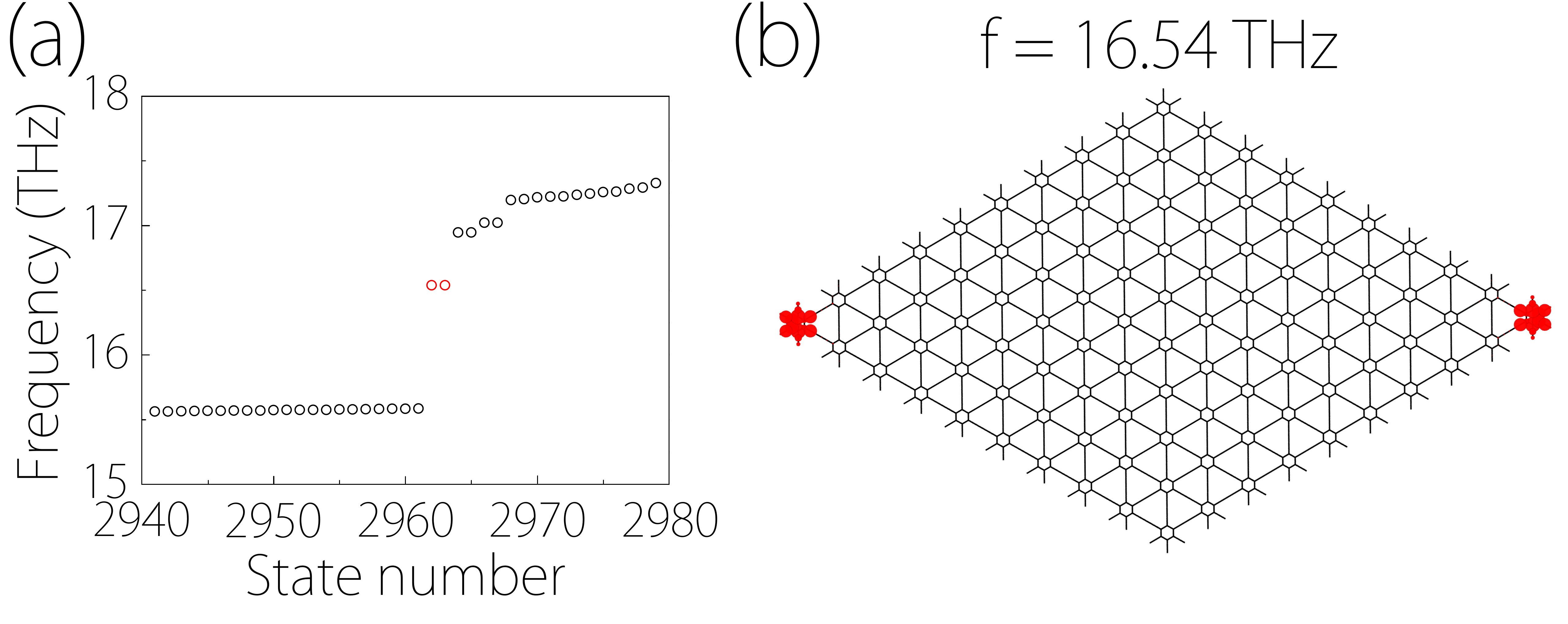}
    \caption{(a) Phonon spectrum for a diamond shaped graphdiyne nanodisk in (b). Again, we focus on the frequency range close to Gap 2. (b) shows the distribution of the two red colored modes at 16.54 THz, which are localized at corners.}
    \label{fig:3}
\end{figure}

In the above calculation, we have taken a hexagonal shaped disk geometry. This is the natural sample geometry that is obtained from the bottom-up synthesis approach~\cite{bottom-up}. Hence, our results here can be directly tested in experiment. Moreover, it must be emphasized that the existence of the corner modes is general and not depend on the specific sample geometry. As demonstrated in Ref.~\cite{2d-ehoti-2}, such modes must exist when the sample geometry preserves $\mathcal{PT}$. In Fig.~\ref{fig:3}, we show the result for a diamond shaped sample. Again, localized corner modes in Gap 2 can be observed.

%
%

\section{Discussion and Conclusion}\label{gyn}

We have demonstrated the phononic HOTI state in graphdiyne. More specifically, it is the real Chern insulator state. In the bulk, there exist phononic band gaps that are characterized by nontrivial real Chern numbers. At the boundary, the nontrivial invariant dictates the existence of corner phonon modes.

\begin{figure}[ht!]
    \centering
    \includegraphics[width=1.0\linewidth]{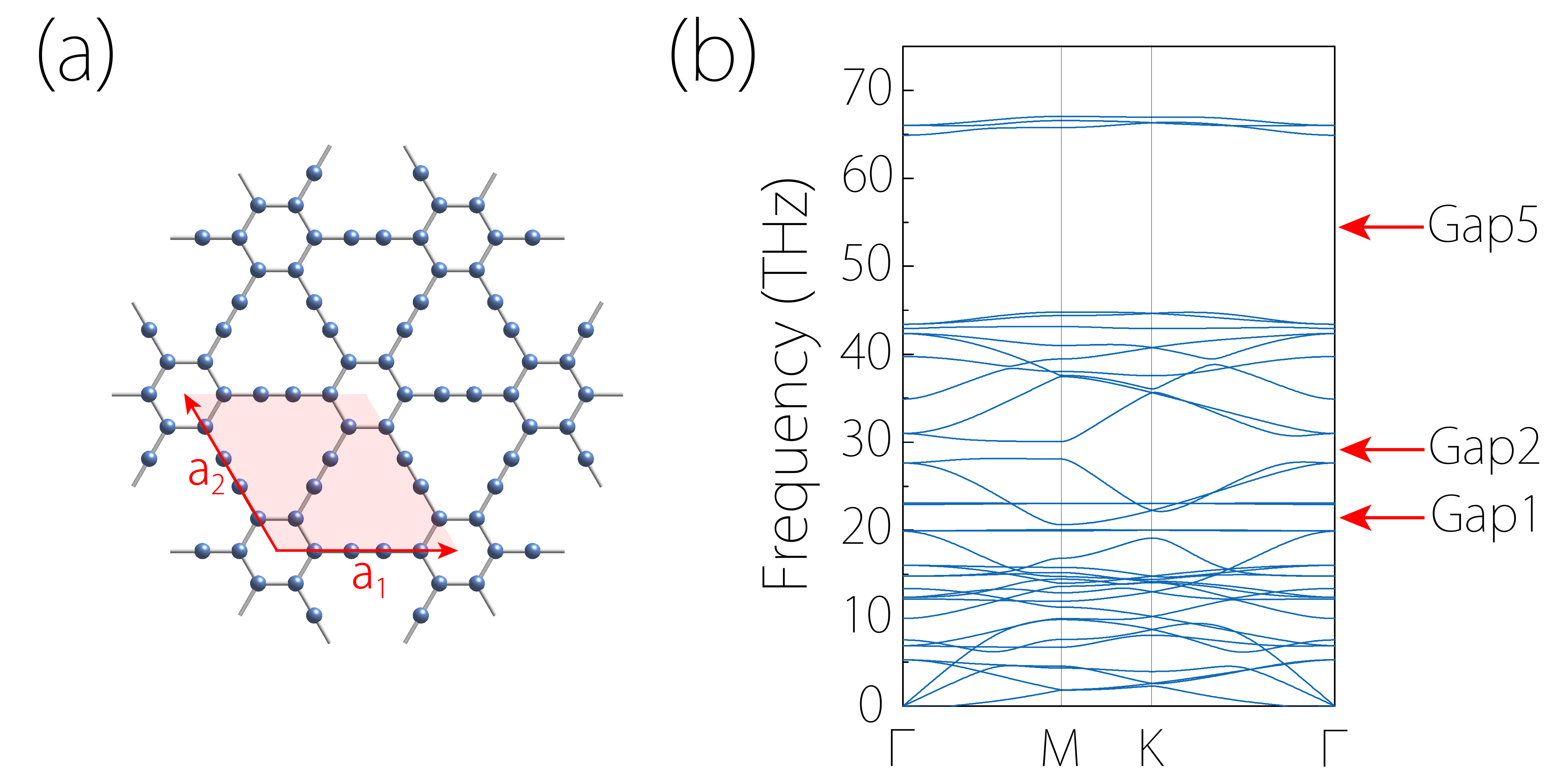}
    \caption{(a) Crystal structure of $\gamma$-graphyne. The unit cell is marked by the shaded region. (b) Phonon spectrum of $\gamma$-graphyne. The arrows indicate three global gaps.}
    \label{fig:4}
\end{figure}

\begin{figure}[ht!]
    \centering
    \includegraphics[width=1.0\linewidth]{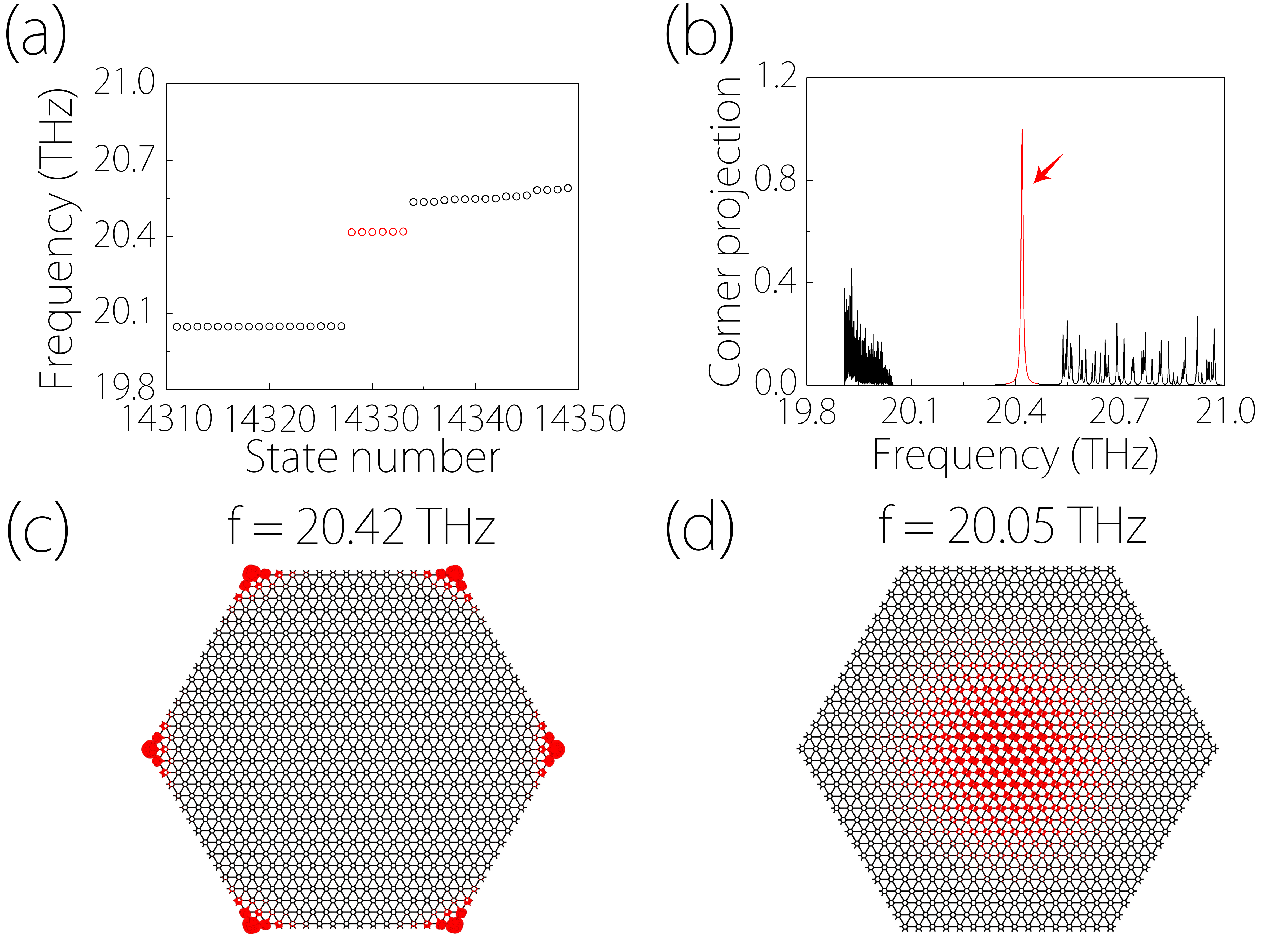}
    \caption{(a) Phonon spectrum for the $\gamma$-graphyne nanodisk shown in (c). Here, we focus on the frequency range close to Gap 1 in Fig.~4(b). (b) Projection spectra at one corner of the disk. The arrow indicate the peak due to the corner modes. (c) Distribution for the red-colored phonon modes at 20.42 THz in (a). For comparison, (d) shows the distribution for bulk modes at 20.05 THz.}
    \label{fig:5}
\end{figure}

The results for $\gamma$-graphyne is similar. The structure of $\gamma$-graphyne is shown in Fig.~\ref{fig:4}(a), where the benzene rings are connected by acetylenic rather than diacetylenic bond. There are 5 phononic band gaps in its phonon spectrum, of which Gap 1 and Gap 5 are identified to be nontrivial and feature $\nu_R=1$ (see Supplemental Material~\cite{sup} for more details). Here, let's consider Gap 1. In Fig.~\ref{fig:5}(c), we show the corner phonon modes correspond to the gap around 20.42 THz for a graphyne nanodisk. Clearly, our approach can be applied to study other members of the graphyne family as well, as all typical members of the family possess both the $\mathcal{P}$ and $\mathcal{T}$ symmetries. Particularly, we expect that the phononic real Chern insulator states can appear in the graphyne-$n$ structures~\cite{carbon-2, carbon-3}, which are generalizations of $\gamma$-graphyne and graphdiyne with extended linkage containing $n$ $-$C$\equiv $C$-$ bonds.


Experimentally, graphdiyne has been successfully synthesized in experiment~\cite{gdy-synthe, carbon-3, carbon-2} and samples with the hexagonal disk geometry can be naturally realized via the bottom-up growth method~\cite{bottom-up}. To probe the corner phonon modes, one can excite them with an infrared light in resonance with their frequency. As the corner modes are spatially localized at the corners, they can manifest in spatially resolved fluorescence pattern, or they can
be detected
via local probes such as scanning tunneling microscopy, by comparing the measurements at the center  and at the corner of the disk.

The localized phonon mode in a spectral gap could be useful for studying many interesting effects. For example, it may offer a new platform to realize the phonon laser effect, which was previously demonstrated in trapped ions/nanosphere and micro-cavities~\cite{phonon-laser-1, phonon-laser-2, phonon-laser-3, phonon-laser-4}. The effect could be useful for constructing novel devices for ultra-sensitive force sensing and high-precision imaging applications.

In conclusion, we have reported the phononic HOTI state, more specifically, the real Chern insulator state, in the graphyne family materials.
Taking graphdiyne and $\gamma$-graphyne as examples, we show that their phonon spectra have nontrivial phononic band gaps characterized by the real Chern numbers enabled by the $\mathcal{PT}$ symmetry. We evaluate the real Chern numbers by the parity analysis approach. The topological corner phonon modes associated with the nontrivial gaps are explicitly demonstrated via calculation on a disk geometry. Our study extends the scope of higher-order topology to phonons in real material systems. The spatially localized corner phonon modes could be useful for novel phononic applications and phonon-photon coupling at nanoscale.

\begin{acknowledgments}
The authors thank D. L. Deng for helpful discussions. This work is supported by Singapore Ministry of Education AcRF Tier 2 (MOE2019-T2-1-001). We acknowledge computational support from the Texas Advanced Computing Center and the National Supercomputing Centre Singapore.
\end{acknowledgments}

\bibliographystyle{apsrev4-1}
\bibliography{ref}

\end{document}